%% file: paper.tex
\pdfoutput=1
\documentclass[10pt,twocolumn]{confpaper}

\graphicspath{ {./figures/} }

\usepackage{cancel}
\usepackage{longtable}
\usepackage[normalem]{ulem}
\usepackage{algorithm}
\usepackage[noend]{algpseudocode}
\usepackage{adjustbox}
\usepackage{array}
\usepackage{dcolumn}
\usepackage{subcaption}
\usepackage{siunitx}
\usepackage[utf8]{inputenc}
\usepackage[english]{babel}
\def\checkmark{\tikz\fill[scale=0.4](0,.35) -- (.25,0) -- (1,.7) -- (.25,.15) -- cycle;} 
\newtheorem{finding}{Finding}[section]

\date{}
\title{\ttlfnt{
Characterizing and Avoiding Routing Detours \\Through Surveillance
States}}
\author{}
\author{
\aufnt{Anne Edmundson, Roya Ensafi, Nick Feamster, Jennifer Rexford} \\
\affaddr{Princeton University}
}

\begin{document}

\maketitle


\begin{sloppypar}

\thispagestyle{empty}
\input{abstract}

\ifthenelse{\equal{\onlyAbstract}{no}}{

\input{intro}
\input{surveillance}

\input{datasets}

\input{avoid_results}
\input{discussion}
\input{related_work}

\input{conclusion}
\label{lastpage}

\end{sloppypar}

\small
\setlength{\parskip}{-1pt}
\setlength{\itemsep}{-1pt}
\balance
\flushleft{\footnotesize\bibliography{paper}}
\bibliographystyle{abbrv}
}{
}

\end{document}

%% file: abstract.tex
\begin{abstract}
An increasing number of countries are passing laws that facilitate the
mass surveillance of Internet traffic. In response, governments and
citizens are increasingly paying attention to the countries that their
Internet traffic traverses. In some cases, countries are taking extreme
steps, such as building new Internet Exchange Points (IXPs), which allow networks to interconnect 
directly, and encouraging local interconnection
to keep local traffic local. We find that although many of these efforts
are extensive, they are often futile, due to the inherent lack of
hosting and route diversity for many popular sites. By measuring the country-level paths to 
popular domains, we characterize transnational routing detours.  We find that traffic is traversing known surveillance 
states, even when the traffic originates and ends in a country that does not conduct mass surveillance.  Then, we investigate how
clients can use overlay network relays and the open DNS resolver
infrastructure to prevent their traffic from traversing certain jurisdictions. We find
 that 84\% of paths originating in Brazil traverse the United States, 
but when relays are used for country avoidance, only 37\% of Brazilian paths 
traverse the United States.  Using the open DNS resolver infrastructure allows Kenyan clients 
to avoid the United States on 17\% more paths.  Unfortunately, we find that 
some of the more prominent surveillance states (e.g., the U.S.) are also some of the least avoidable 
countries.
\end{abstract}

%% file: intro.tex
\section{Introduction}
\label{intro}

When Internet traffic enters a country, it becomes subject to that
country's laws.  As a result, users have more need than ever to
determine---and control---which countries their traffic is traversing.
An increasing number of countries have passed laws that facilitate mass
surveillance of their networks~\cite{france_surveillance,
  netherlands_surveillance, kazak_surveillance, uk_bill}, and governments
and citizens are increasingly motivated to divert their Internet traffic
from countries that perform surveillance (notably, the United
States~\cite{russia_secure_internet,
  routing_errors, dte}).

Many countries---notably, Brazil---are taking impressive measures to
reduce the likelihood that Internet traffic transits the United
States~\cite{brazil_history, brazil_break_from_US, brazil_conference,
  brazil_conference2, brazil_human_rights} including building a
3,500-mile long fiber-optic cable from Fortaleza to Portugal (with no
use of American vendors); pressing companies such as Google, Facebook,
and Twitter (among others) to store data locally; and mandating the
deployment of a state-developed email system (Expresso) throughout the
federal government (instead of what was originally used, Microsoft
Outlook)~\cite{brazil_cable, brazil_us_companies}.  Brazil is also
building Internet Exchange Points (IXPs)~\cite{brazil_IXP1}, now has the
largest national ecosystem of public IXPs in the
world~\cite{brazil_ixp_ecosystem}, and the number of internationally
connected Autonomous Systems (ASes) continues to
grow~\cite{brazil_international_ases}. Brazil is not alone: IXPs are
proliferating in eastern Europe, Africa, and other regions, in part out
of a desire to ``keep local traffic local''. Building IXPs alone, of
course, cannot guarantee that Internet traffic for some service does not
enter or transit a particular country: Internet protocols have no notion
of national borders, and interdomain paths depend in large part on
existing interconnection business relationships (or lack thereof).

Although end-to-end encryption stymies surveillance by
concealing URLs and content, it does not by itself protect all sensitive
information from prying eyes. First, many websites do not fully support
encrypted browsing by default; a recent study showed that more than 85\%
of the most popular health, news, and shopping sites do not encrypt by
default~\cite{what_isps_can_see}; migrating a website to HTTPS is
challenging doing so requires all third-party domains on the site
(including advertisers) to use HTTPS.  Second, even encrypted traffic
may still reveal a lot about user behavior: the presence of any
communication at all may be revealing, and website fingerprinting can
reveal information about content merely based on the size, content, and
location of third-party resources that a client loads. DNS traffic is
also quite revealing and is essentially never
encrypted~\cite{what_isps_can_see}.  Third, ISPs often terminate TLS
connections, conducting man-in-the-middle attacks on encrypted traffic
for network management purposes~\cite{mitm_isp}. Circumventing
surveillance thus requires not only encryption, but also mechanisms
for controlling where traffic goes in the first place.

In this paper, we study two questions: (1)~Which countries do {\em
  default} Internet routing paths traverse?; (2)~What methods can help
increase hosting and path diversity to help governments and citizens
better control transnational Internet paths?  In contrast to previous
work~\cite{karlin2009nation}, which simulates Internet paths, we
\textit{actively measure} and analyze the paths originating in five
different countries: Brazil, Netherlands, Kenya, India, and the United
States.  We study these countries for different reasons, including their
efforts made to avoid certain countries, efforts in building out IXPs,
and their low cost of hosting domains.  Our work studies the
router-level forwarding path, which differs from all other work in this
area, which has focused on analyzing Border Gateway Protocol
(BGP) routes~\cite{karlin2009nation,shah2015characterizing}.  Although
BGP routing can offer useful information about paths, it does not
necessarily reflect the path that traffic actually takes, and it only
provides AS-level granularity, which is often too coarse to make strong
statements about which countries that traffic is traversing.  In
contrast, we measure traffic routes from RIPE Atlas probes in five
countries to the Alexa Top 100 domains for each country; we directly
measure the paths not only to the websites corresponding to the
themselves, but also to the sites hosting any third-party content on
each of these sites.

Determining which countries a client's traffic traverses is challenging, for
several reasons.  First, performing direct measurements is more costly
than passive analysis of BGP routing tables; RIPE Atlas, in particular,
limits the rate at which one can perform measurements.  As a result, we
had to be strategic about the origins and destinations that we selected
for our study. As we explain in Section~\ref{surv}, we study five
geographically diverse countries, 
focusing on countries in each region that are
making active attempts to thwart transnational Internet paths.  Second,
IP geolocation---the process of determining the geographic location of an
IP address---is notoriously challenging, particularly for IP addresses
that represent Internet infrastructure, rather than end-hosts. We cope
with this inaccuracy by making conservative estimates of the extent of
routing detours, and by recognizing that our goal is not to pinpoint a
precise location for an IP address as much as to achieve accurate
reports of {\em significant} off-path detours to certain countries or
regions. (Section~\ref{datasets} explains our method in more detail; we
also explicitly highlight ambiguities in our results.) Finally, the
asymmetry of Internet paths can also make it difficult to analyze the
countries that traffic traverses on the reverse path from server to
client; our study finds that country-level paths are often asymmetric,
and, as such, our findings represent a lower bound on transnational
routing detours.

The first part of our study (Section~\ref{datasets}) characterizes the
current state of transnational Internet routing detours.  We first
explore hosting diversity and find that only about half of the Alexa Top
100 domains in the five countries studied are hosted in more than one
country, and many times that country is a surveillance state that
clients may want to avoid. Second, even if hosting diversity can be
improved, routing can still force traffic through a small
collection of countries (often surveillance states). Despite strong
efforts made by some countries to ensure their traffic does not transit
unfavorable countries~\cite{brazil_history, brazil_break_from_US,
  brazil_conference, brazil_conference2, brazil_human_rights}, their
traffic still traverses surveillance states.  Over 50\% of the top
domains in Brazil and India are hosted in the United States, and over
50\% of the paths from the Netherlands to the top domains transit the
United States.  About half of Kenyan paths to the top domains traverse
the United States and Great Britain (but the same half does not traverse
both countries).  Much of this phenomenon is due to ``tromboning'',
whereby an Internet path starts and ends in a country, yet transits an
intermediate country; for example, about 13\% of the paths that we
explored from Brazil tromboned through the United States.
Infrastructure building alone is not enough: ISPs in respective regions
need better encouragements to interconnect with one another to ensure
that local traffic stays local.

The second part of our work (Section~\ref{avoid_results}) explores
potential mechanisms for avoiding certain countries, and the potential
effectiveness of these techniques.  We explore two techniques: using the
open DNS resolver infrastructure and using overlay network relays.  We
find that both of these techniques can be effective for clients in
certain countries, yet the effectiveness of each technique also depends
on the county.  For example, Brazilian clients could completely avoid
Spain, Italy, France, Great Britain, Argentina, and Ireland (among
others), even though the default paths to many popular Brazilian sites
traverse these countries. Additionally, overlay network relays can keep
local traffic local: by using relays in the client's country, fewer
paths trombone out of the client's country.  The percentage of
tromboning paths from the United States decreases from 11.2\% to 1.3\%
when clients take advantage of a small number of overlay network relays.

We also find that some of the most prominent surveillance states are
also some of the least avoidable countries.  For example, many countries
depend on ISPs in the United States, a known surveillance state, for
connectivity to popular sites and content. Brazil, India, Kenya, and
the Netherlands must traverse the United States to reach many of the
popular local websites, even if they use open resolvers and network
relays. Using overlay network relays, both Brazilian and Netherlands
clients can avoid the United States for about 65\% of paths; yet, the
United States is completely unavoidable for about 10\% of the paths
because it is the only country where the content is hosted.  Kenyan
clients can only avoid the United States on about 55\% of the paths.  On
the other hand, the United States can avoid every other country except
for France and the Netherlands, and even then they are avoidable for
99\% of the top domains.

%% file: surveillance.tex
\section{State of Surveillance}
\label{surv}
We focused our study on five different countries, and for each, we actively measured and analyzed traffic that originated there.  These five countries were chosen for specific reasons and we present them here.  We also discuss countries that currently conduct surveillance; this exploration is not exhaustive, but highlights countries that are passing new surveillance laws and countries that have strict surveillance practices already.   

\subsection{Studied Countries}
We selected Brazil, Netherlands, Kenya, India, and the United States for the following reasons.

\paragraph{Brazil.} It has been widely publicized that Brazil is actively trying to avoid having their traffic transit the United States.  They have been building IXPs, deploying underwater cables to Europe, and pressuring large U.S. companies to host content within Brazil~\cite{brazil_history, brazil_break_from_US, brazil_conference,
  brazil_conference2, brazil_human_rights, brazil_cable, brazil_us_companies, brazil_IXP1}.  This effort to avoid traffic transitting a specific country led us to investigate whether their efforts have been successful or not.

\paragraph{Netherlands.}  We selected to study the Netherlands for three reasons: 1) the Netherlands is beginning to emerge as a site where servers are located for cloud services, such as Akamai, 2) the Netherlands is where a large IXP is located (AMS-IX), and 3) they are drafting a mass surveillance law~\cite{netherlands_surveillance}. Analyzing the Netherlands will allow us to see what effect AMS-IX and the emergence of cloud service hosting has on their traffic.

\paragraph{Kenya.} Prior research on the interconnectivity of Africa~\cite{gupta2014peering, fanou2015diversity} led us to explore the characterization of an African country's interconnectivity.  We chose Kenya for a few reasons: 1) it is a location with many submarine cable landing points, 2) it has high Internet access and usage (for the East African region), and 3) it has more than one IXP~\cite{kenya_nigeria, teams}.

\paragraph{India.}  India has one of the highest number of Internet users in Asia, second only to China, which has already been well-studied~\cite{tsui2003panopticon, wang2010discourse}.  With such a high number of Internet users, and presumably a large amount of Internet traffic, we study India to see where this traffic is going.

\paragraph{United States.}  We chose to study the United States because of how inexpensive it is to host domains there, the prevalence of Internet and technology companies located there, and because it is a known surveillance state.

\subsection{Surveillance States}

When analyzing which countries Internet traffic traverse, special attention should be given to countries that may be unfavorable because of their surveillance laws.  Some of the countries that are currently conducting surveillance are the ``Five Eyes'' ~\cite{lander2004international, eyeswideopen} (the United States, Canada, United Kingdom, New Zealand, and Australia), as well as France, Germany, Poland, Hungary, Russia, Ukraine, Belarus, Kyrgyzstan, and Kazakhstan.  

\paragraph{Five Eyes.} The ``Five Eyes'' participants are the United States National Security Agency (NSA), the United Kingdom's Government Communications Headquarters (GCHQ), Canada's Communications Security Establishment Canada (CSEC), the Australian Signals Directorate (ASD), and New Zealand's Government Communications Security Bureau (GCSB)~\cite{eyeswideopen}.  According to the original agreement, the agencies can: 1) collect traffic; 2) acquire communications documents and equipment; 3) conduct traffic analysis; 4) conduct cryptanalysis; 5) decrypt and translate; 6) acquire information about communications organizations, procedures, practices, and equipment.  The agreement also implies that all five countries will share all intercepted material by default.  The agencies work so closely that the facilities are often jointly staffed by members of the different agencies, and it was reported ``that SIGINT customers in both capitals seldom know which country generated either the access or the product itself.''~\cite{lander2004international}.

\begin{table}[t!]
\centering
\begin{small}
\resizebox{\columnwidth}{!}{%
\begin{tabular}{|p{2cm}|p{1.6cm}p{1.6cm}p{1.6cm}p{1.6cm}|}
\hline
 & Collecting Metadata (Phone, Internet) & Requiring ISPs to Participate & No Need for Court Order & Targeted Surveillance \\
\hline
France            & \checkmark~\cite{francesurv, francesurv2} & \checkmark~\cite{francesurv} &  & \\ 
Germany           & \checkmark~\cite{germansurv}   &             &             &                  \\ 
UA Emirates & & & & \checkmark~\cite{uae_surv} \\
Bahrain & & & & \checkmark~\cite{bahrain_surv} \\
Australia & \checkmark~\cite{eyeswideopen} & &  &\\
New Zealand & \checkmark~\cite{eyeswideopen} & &  & \\
Canada & \checkmark~\cite{eyeswideopen} & & & \\
United States & \checkmark~\cite{eyeswideopen} & & & \\
Great Britain & \checkmark~\cite{eyeswideopen} & & & \\
Poland            & \checkmark~\cite{francesurv2}      &         &  \checkmark~\cite{francesurv2}      &      \\ 
Hungary            & \checkmark~\cite{francesurv2}        &             & \checkmark~\cite{francesurv2} & \\ 
Ukraine            & \checkmark~\cite{francesurv2}    & \checkmark~\cite{russiasurv, russiasurv2}    & &  \\ 
Belarus            & \checkmark~\cite{francesurv2}    & \checkmark~\cite{russiasurv, russiasurv2}    & &  \\ 
Kyrgyzstan            & \checkmark~\cite{francesurv2}    & \checkmark~\cite{russiasurv, russiasurv2}    & &  \\ 
Kazakhstan            & \checkmark~\cite{francesurv2}    & \checkmark~\cite{russiasurv, russiasurv2}    & &  \\ 
Russia            & \checkmark~\cite{francesurv2}    & \checkmark~\cite{russiasurv, russiasurv2}    & &  \\ \hline
\end{tabular}
}
\end{small}
\caption{Some countries that actively conduct surveillance.}
\label{surv_table}
\end{table}

A number of other countries are passing laws to facilitate mass surveillance.  These laws have differing levels of intensity, which can be seen in Table \ref{surv_table}; the countries with the least intense surveillance laws are listed at the top of the table, and those with the more intense laws are listed at the bottom.  These countries, along with the ``Five Eyes'' participants should be flagged when characterizing transnational detours in the following section.

%% file: datasets.tex
\section{Characterizing Transnational Detours}
\label{datasets}
In this section, we describe our measurement methods, the challenges in
conducting them, and our findings concerning the transnational detours
of default Internet paths.

\begin{figure*}[t]
\centering
\includegraphics[width=.9\textwidth]{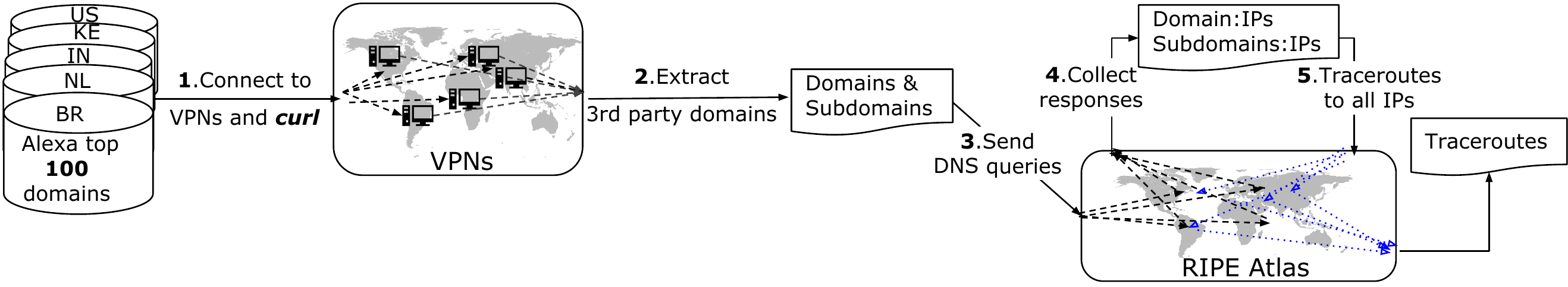}
\caption{Measurement pipeline to study Internet paths from countries to
  popular domains.}
\label{fig:pipeline1}
\end{figure*}

\subsection{Measurement Pipeline}
\label{pipeline}

Figure \ref{fig:analysis_pipeline} summarizes our measurement
process, which the rest of this section describes in detail.  We analyze
traceroute measurements to discover which countries are on the path from
a client in a particular country to a popular domain.  Using traceroutes
to measure transnational detours is new; prior work used BGP routing
tables to \textit{infer} country-level paths~\cite{karlin2009nation}.
Because we conduct active measurements, which are limited by our
resources, we make a tradeoff and study five countries, as opposed to
all countries' Internet paths.  We report on measurements that we
conducted on January 31, 2016.

\subsubsection{Resource Limitations}
\label{resource_limits}

The iPlane~\cite{madhyastha2006iplane} and Center for Applied
Internet Data Analysis (CAIDA)~\cite{caida} projects maintain two large
repositories of traceroute data, neither of which turn out to be
suitable for our study.  iPlane measurements use
PlanetLab~\cite{planetlab} nodes and has historical data as far back as
2006.  Unfortunately, because iPlane uses PlanetLab nodes, which 
mostly use the Global Research and Education Network
(GREN), the traceroutes from PlanetLab nodes will not be representative
of typical Internet users' traffic paths~\cite{banerjee2004interdomain}.
CAIDA runs traceroutes from different vantage points around the world to
randomized destination IP addresses that cover all /24s; in 
contrast, we focus on paths to popular websites from a particular
country.

In contrast to these existing studies, we run active measurements that
would represent paths of a typical Internet user. To do so, we run
DNS and traceroute measurements from RIPE Atlas probes, which are hosted
all around the world and in many different settings, including home
networks~\cite{ripe_atlas}.  RIPE Atlas probes can use the local DNS
resolver, which would give us the best estimate of the traceroute
destination.

Yet, conducting measurements from a RIPE Atlas probe costs a certain
amount of ``credits'', which restricts the number of measurements that we
could run.  RIPE Atlas also imposes rate limits on the number of
concurrent measurements and the number of credits that an individual
user can spend per day.  We address these challenges in two ways: (1)~we
reduce the number of necessary measurements we must run on RIPE Atlas
probes by conducting traceroute measurements to a single IP address in
each /24 (as opposed to all IP address returned by DNS) because all IP
addresses in a /24 belong to the same AS, and should therefore be
located in the same geographic area; (2)~we use a different method---VPN
connections---to obtain a vantage point within a foreign country, which
is still representative of an Internet user in that country.

\subsubsection{Path Asymmetry}
\label{path_sym}

The reverse path is just as important as (and often different from) the 
forward path.  
Previous work has shown that paths are not symmetric most of the
time---the forward path from point A to point B does not match the
reverse path from point B to point A~\cite{he2005routing}.  Most work on
path asymmetry has been done at the AS level, but not at the country
level.  Our measurements consider only the forward path (from client
to domain or relay), not the reverse path from the domain or
relay to the client.   

We measured path asymmetry at the country
granularity. If country-level paths are symmetric, then the results of
our measurements would be representative of the forward {\it and}
reverse paths. If the country-level paths are asymmetric, then our
measurement results only provide a lower bound on the number of countries
that could potentially conduct surveillance.  Using 100 RIPE Atlas
probes located around the world, and eight Amazon EC2 instances, we ran
traceroute measurements from every probe to every EC2 instance and from
every EC2 instance to every probe.  After mapping the IPs to countries,
we analyzed the paths for symmetry.  First, we compared the set of
countries on the forward path to the set of countries on the reverse
path; this yielded about 30\% symmetry.  What we wanted to know is
whether or not the reverse path has more countries on it than the
forward path.  Thus, we measured how many reverse paths were a subset of the
respective forward path; this was the case for 55\% of the paths.   
This level of asymmetry suggests that our results represent a lower
bound on the number of countries that transit traffic; our results are a
lower bound on how many unfavorable countries transit a client's
path. It also suggests that while providing lower bounds on
transnational detours is feasible, designing systems to completely
prevent these detours on both forward and reverse paths may be particularly
challenging, if not impossible. 

\subsubsection{Traceroute Origination and Destination Selection}

Each country hosts a different number of RIPE Atlas probes, ranging
from about 75 probes to many hundreds.  Because of the resource
restrictions, we could not use all probes in each of the countries.  We
selected the set of probes that had unique ASes in the country to get
the widest representation of origination (starting) points.

For destinations, we used the Alexa Top 100 domains in each of the
respective countries, as well as the third-party domains that are
requested as part of an original web request.  To obtain these 3rd party
domains we {\tt curl} (\ie, HTTP fetch) each of the Top 100 domains, but we must
do so from within the country we are studying.  There is no current
functionality to {\tt curl} from RIPE Atlas probes, so we establish a
VPN connection within each of these countries to {\tt curl} each domain
and extract the third-party domains; we {\tt curl} from the client's
location in case web sites are customizing content based on the region
of the client. 

\begin{figure}[t]
\centering
\includegraphics[width=.5\textwidth]{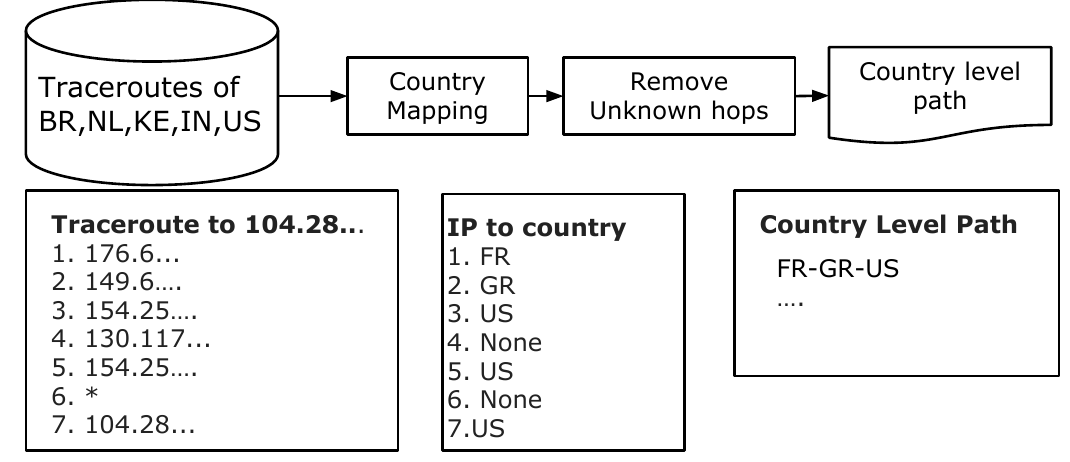}
\caption{Mapping country-level paths from traceroutes.}
\label{fig:analysis_pipeline}
\end{figure}

\subsubsection{Country Mapping}
\label{c_map}

Accurate IP geolocation is challenging. We use MaxMind's
geolocation service to map IP addresses to their respective
countries~\cite{maxmind}, which is known to contain inaccuracies.
Fortunately, our study does not require high-precision geolocation; we
are more interested in providing accurate lower bounds on detours at a
much coarser granularity.  Fortunately, previous work has found that
geolocation at a country-level granularity is more accurate than at
finer granularity~\cite{huffaker2011geocompare}.  In light of these
concerns, we post-processed our IP to country mapping
by removing all IP addresses that resulted in a `None' response when
querying MaxMind, which causes our results to provide a conservative
estimate of the number of countries that paths traverse. It is important
to note that removing `None' responses will \textit{always} produce a
conservative estimate, and therefore we are \textit{always}
underestimating the amount of potential surveillance.  
Figure \ref{fig:analysis_pipeline} shows an example of this
post-processing. 

\subsection{Results}

\input{characterize-tables}

Table \ref{tab:host} shows the five countries we studied along the top
of the table, and the countries that host their content along in each
row.  For example, the United States is the endpoint of 77\% of the
paths that originate in Brazil.  A ``-'' represents the case where no
paths ended in that country.  For example, no Brazilian paths terminated in
South Africa. Table \ref{tab:transit} shows the fraction of paths that
transit certain countries, with a row for each country that is transited.

\begin{figure}[t]
\centering
\includegraphics[width=\columnwidth]{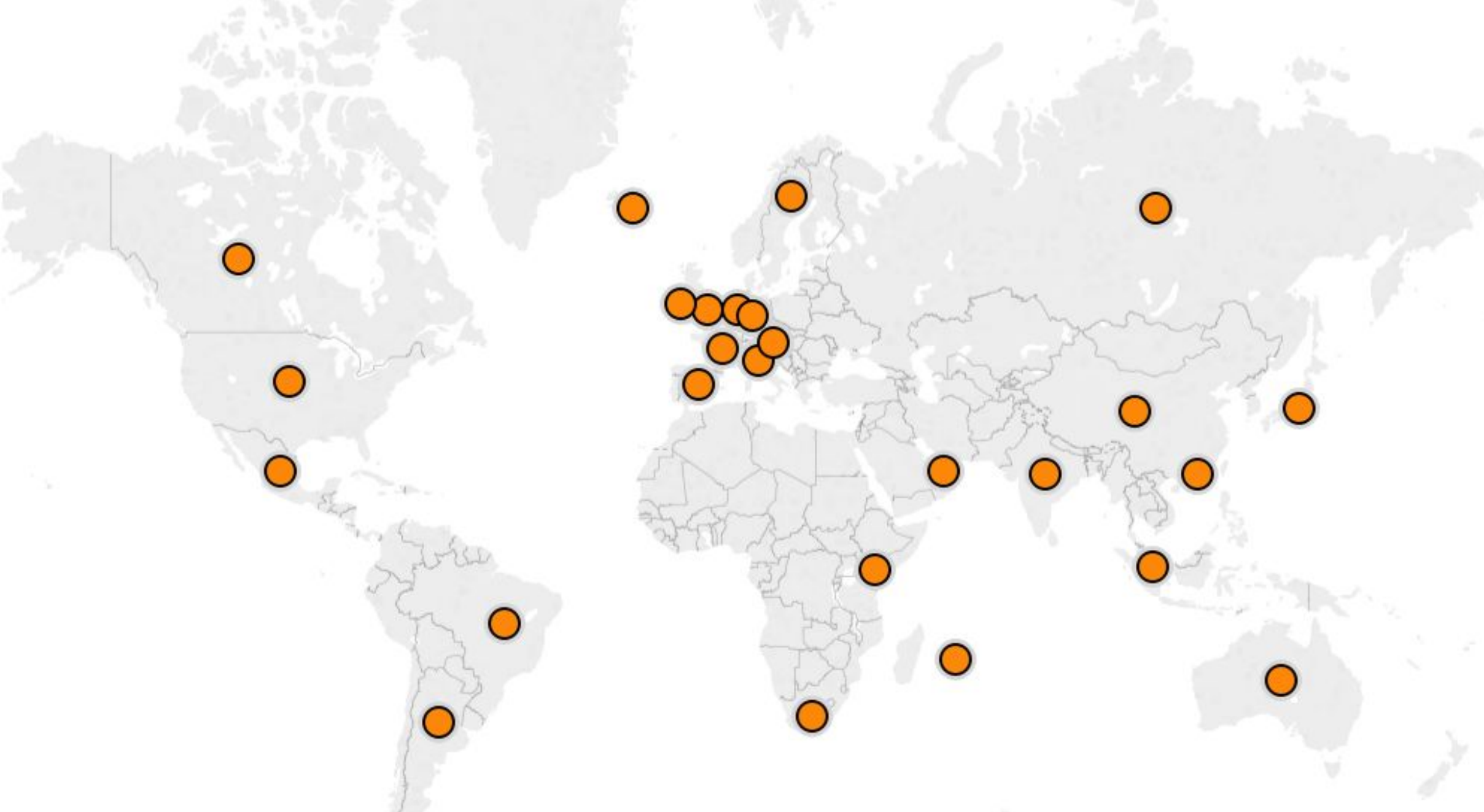}
\caption{The locations of vantage points in measuring hosting diversity.}
\label{fig:world}
\end{figure}

\begin{figure}[t]
\centering
\includegraphics[width=\columnwidth]{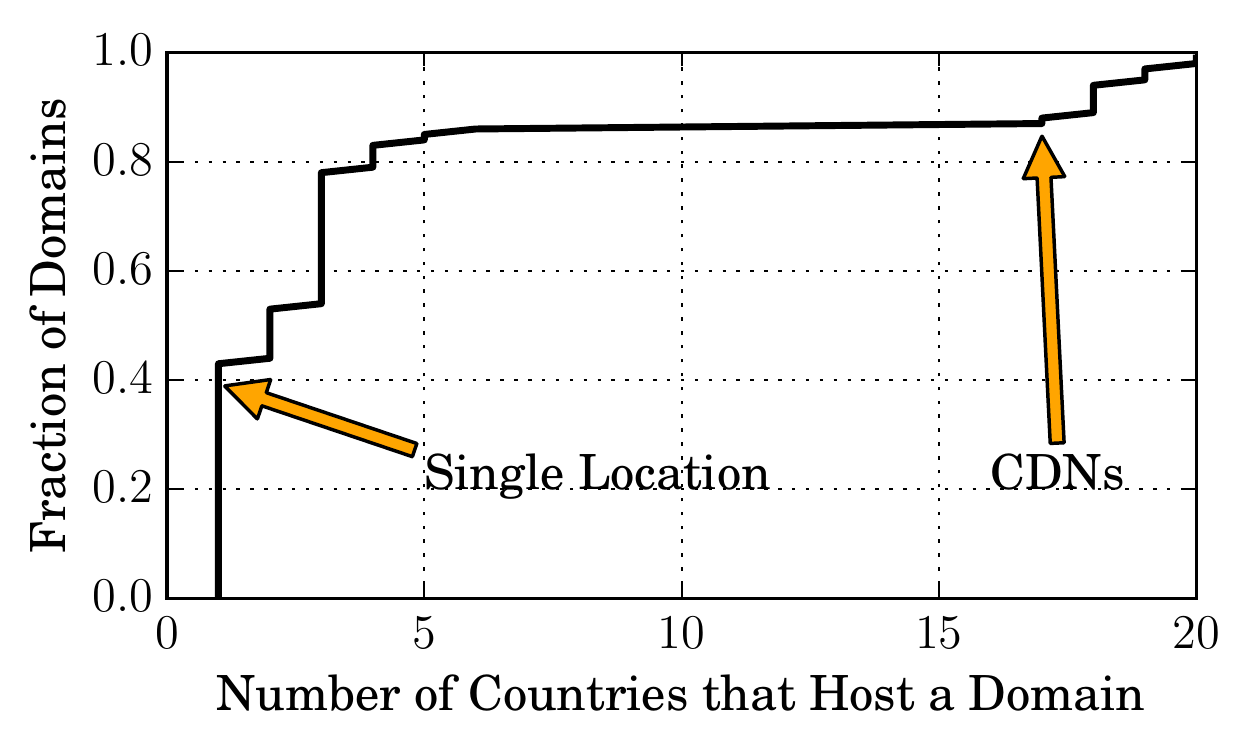}
\caption{The number of Alexa Top 100 US Domains hosted in different countries.}
\label{fig:host_diversity}
\end{figure}

\begin{finding}[Hosting Diversity]
About half of the top domains in each of the five countries studied are hosted in a single country.  The other half are located in two or more different countries.
\end{finding}
\noindent
First we analyze hosting diversity; this shows us how many unique
countries host a domain.  The more countries that a domain is hosted in creates a greater chance that the content is replicated in a favorable country, and could potentially allow a client to circumvent an unfavorable country.  We queried DNS from 26 vantage points around the world, which are shown in Figure \ref{fig:world}; we chose this set of locations because they are geographically diverse.  Then we mapped the IP addresses in the DNS responses to their respective countries to determine how many unique countries a domain is hosted in.  Figure~\ref{fig:host_diversity} shows the fraction of domains that are hosted in different numbers of countries; we can see two common hosting cases: 1) CDNs, and 2) a single hosting country.  This shows that many domains are hosted in a single unique country, which leads us to our next analysis---where are these domains hosted, and which countries are traversed on the way to reach these locations.

\begin{figure*}[t!]
\begin{minipage}{\linewidth}
\begin{subfigure}[b]{.32\linewidth}
\includegraphics[width=\linewidth]{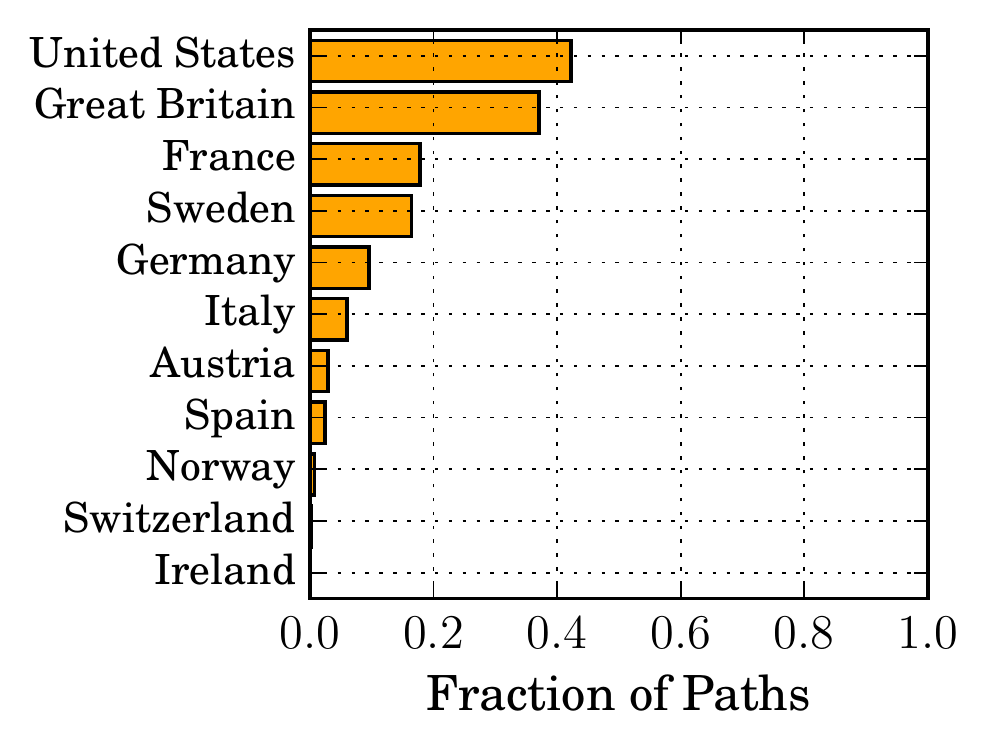}
\caption{The Netherlands.\label{fig:trombone_netherlands}}
\end{subfigure}
\begin{subfigure}[b]{.32\linewidth}
\includegraphics[width=\linewidth]{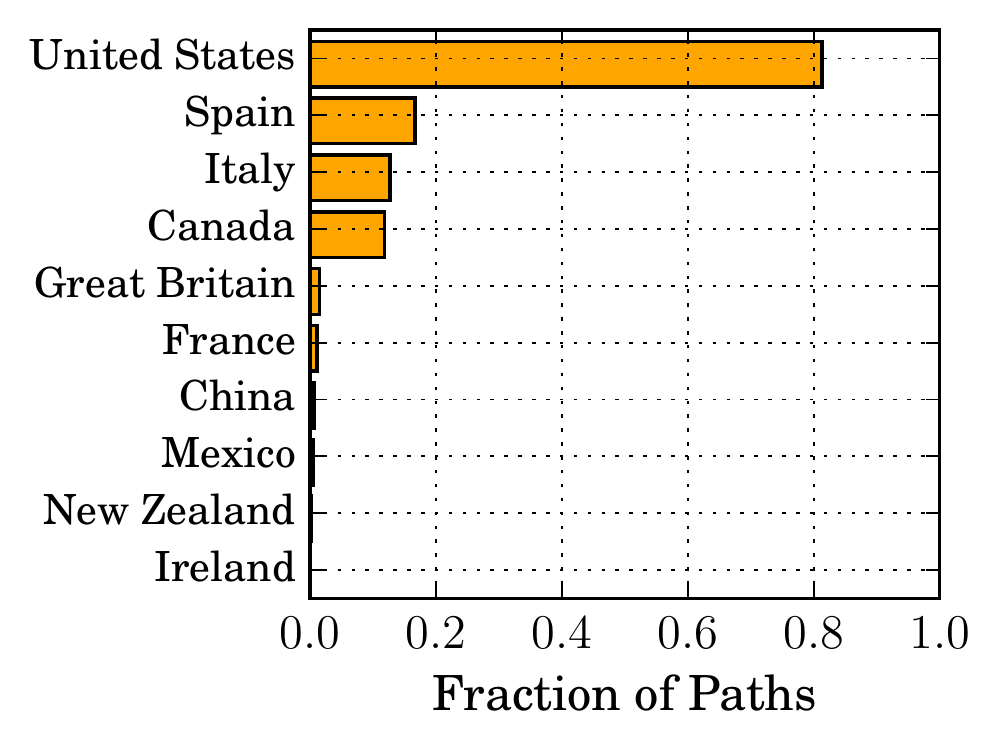}
\caption{Brazil.\label{fig:trombone_brazil}}
\end{subfigure}
\begin{subfigure}[b]{.32\linewidth}
\includegraphics[width=\linewidth]{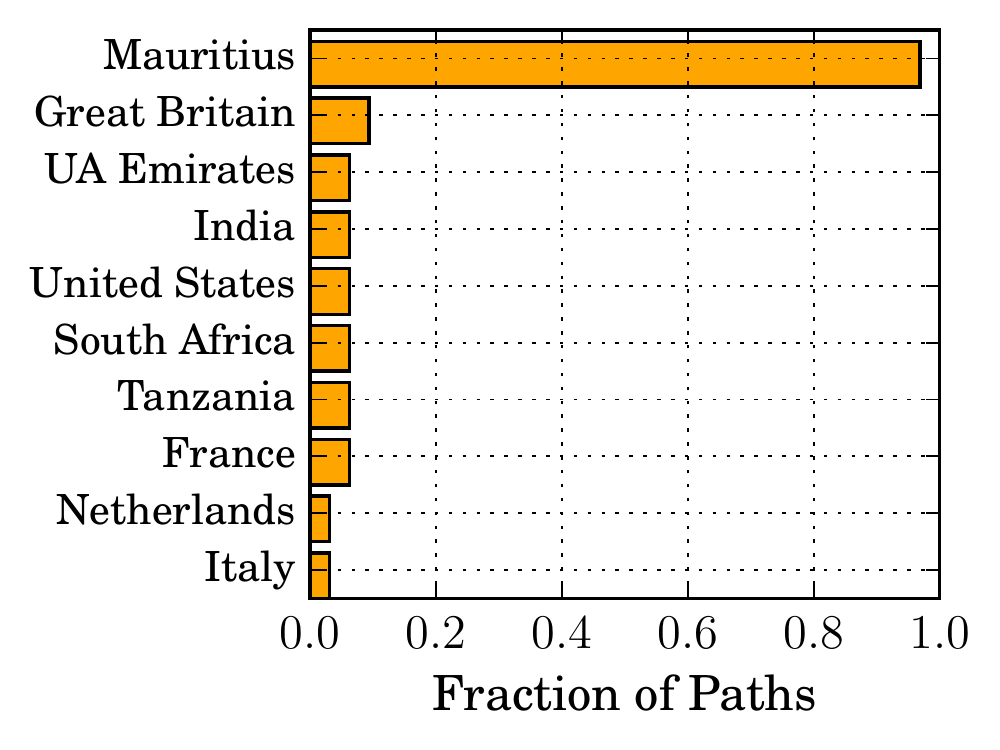}
\caption{Kenya.\label{fig:trombone_kenya}}
\end{subfigure}
\end{minipage}
\caption{The countries that tromboning paths from the Netherlands, Brazil, and Kenya transit.}
\label{fig:trombone}
\end{figure*}

\begin{finding}[Domain Hosting]
The most common destination among all five countries studied is the United States: 77\%, 45\%, 63\%, 44\%, and 97\% of paths originating in Brazil, Netherlands, India, Kenya and the United States, respectively, are currently reaching content located in the United States.
\end{finding}
\noindent
Table~\ref{tab:host} shows the fraction of paths that are hosted in
various countries.  Despite the extent of country-level hosting
diversity, the majority of paths from all five countries terminate in a
single country: the United States, a known surveillance state.   Our results also show the Netherlands is a
common hosting location for paths originating in the Netherlands, India,
and Kenya.



\begin{finding}[Domestic Traffic]
All of the countries studied (except for the United States) host content for a small percentage of the paths that originate in their own country; they also host a small percentage of their respective country-code top-level domains.
\end{finding}
\noindent
Only 17\% of paths that originate in Brazil also end there.  Only 5\%
and 2\% of Indian and Kenyan paths, respectively, end in the originating
country.  
For Kenya, 24 out of the Top 100 Domains are .ke domains, and of these 24 domains only 5 are hosted within Kenya.  29 out of 40 .nl domains are hosted in the Netherlands; 4 of 13 .in domains are hosted in India; 18 of 39 .br domains are hosted in Brazil.  Interestingly, all .gov domains were hosted in their respective country. 

\begin{finding}[Transit Traffic]
Surveillance states (specifically the United States and Great Britain) are on the largest portion of paths in comparison to any other (foreign) country.
\end{finding}
\noindent
84\% of Brazilian paths traverse the United States, despite Brazil's
strong efforts to avoid United States surveillance.  Although India and
Kenya are geographically distant, 72\% and 62\% of their paths also transit
the United States.

Great Britain and the Netherlands are on the path for a significant
percentage of paths originating in India and Kenya: 50\% and 20\% of
paths that originate in Kenya and India, respectively, transit Great
Britain.   Many paths likely traverse Great Britain and the Netherlands due to
the presence of large Internet Exchange Points (\ie, LINX, AMS-IX).
Mauritius, South Africa, and the United Arab Emirates transit 32\%,
33\%, and 15\% of paths from Kenya.  There are direct underwater cables
from Kenya to Mauritius, and from Mauritius to South
Africa~\cite{cablemap}.  Additionally, there is a cable from Mombasa,
Kenya to Fujairah, United Arab Emirates, which likely explains the large
fraction of paths that include these countries.

\begin{finding}[Tromboning Traffic]
Brazilian and Netherlands paths often trombone to the United States, despite the prevalence of IXPs in both countries.
\end{finding}
\noindent
Figures \ref{fig:trombone_netherlands}, \ref{fig:trombone_brazil}, and \ref{fig:trombone_kenya} show the fraction of paths that trombone to different countries for the Netherlands, Brazil, and Kenya. 24\% of all paths originating in the Netherlands (62\% of domestic paths) trombone to a foreign country before returning to the Netherlands. Despite Brazil's strong efforts in building IXPs to keep local traffic local, we can see that their paths still trombone to the United States.  This is due to IXPs being seen as a threat by competing commercial providers; providers are sometimes concerned that ``interconnection'' will result in making business cheaper for competitors and stealing of customers~\cite{ixp_policy}.  It is likely that Brazilian providers see other Brazilian providers as competitors and therefore as a threat at IXPs, which cause them to peer with international providers instead of other local providers.  Additionally, we see Brazilian paths trombone to Spain and Italy. We have observed that MaxMind sometimes mislabels IP addresses to be in Spain when they are actually located in Portugal.  This mislabelling does not affect our analysis of detours through surveillance states, as we do not highlight either Spain or Portugal as a surveillance state.  We see Italy often in tromboning paths because Telecom Italia Sparkle is one of the top global Internet providers~\cite{bakers}.

Tromboning Kenyan paths most commonly traverse Mauritius, which is expected considering the submarine cables between Kenya and Mauritius.  Submarine cables also explain South Africa, Tanzania, and the United Arab Emirates on tromboning paths.  

\begin{finding}[United States as an Outlier]
The United States hosts 97\% of the content that is accessed from within the country, and only five foreign countries---France, Germany, Ireland, Great Britain, and the Netherlands---host content for the other 3\% of paths.
\end{finding}
\noindent
Many of the results find that Brazilian,
Netherlands, Indian, and Kenyan paths often transit surveillance states,
most notably the United States.  The results from studying paths that
originate in the United States are drastically different from those of
the other four countries.  The other four countries host very small
amounts of content accessed from their own country, whereas the United
States hosts 97\% of the content that is accessed from within the
country.  Only 13 unique countries are ever on a path from the United
States to a domain in the top 100 (or third party domain), whereas 30,
30, 25, and 38 unique countries are seen on the paths originating in
Brazil, Netherlands, India, and Kenya, respectively.   


\subsection{Limitations}

This section discusses the various limitations of our measurement methods
and how they may affect the results that we have reported.

\paragraph{Traceroute accuracy and completeness.}
Our study is limited by the accuracy and completeness of traceroute.
Anomalies can occur in traceroute-based
measurements~\cite{augustin2006avoiding}, but most traceroute anomalies
do not cause an overestimation in surveillance states.  The
incompleteness of traceroutes, where a router does not respond, causes
our results to underestimate the number of surveillance states, and
therefore also provides a lower bound on surveillance. 

\paragraph{IP Geolocation vs.\ country mapping.}
Previous work has shown that there are fundamental challenges in
deducing a geographic location from an IP address, despite using
different methods such as DNS names of the target, network delay
measurements, and host-to-location mapping in conjunction with BGP
prefix information~\cite{padmanabhan2001investigation}.  While it has
been shown that there are inaccuracies and incompleteness in MaxMind's
data~\cite{huffaker2011geocompare}, the focus of this work is on
measuring and avoiding surveillance.  We use Maxmind to map
IP to country (as described in Section \ref{c_map}), which provides a lower
bound on the amount of surveillance, as we have described. 

\paragraph{IPv4 vs.\ IPv6 connectivity.}
The measurements we conducted only collect and analyze IPv4 paths, and
therefore all IPv6 paths are left out of our study.  IPv6 paths likely
differ from IPv4 paths as not all routers that support IPv4 also support
IPv6.  Future work includes studying IPv6 paths and which countries they
transit, as well as a comparison of country avoidability between IPv4
and IPv6 paths.

%% file: characterize-tables.tex
\newcolumntype{d}[1]{D{.}{.}{#1}}
\newcommand{\headrow}[1]{\multicolumn{1}{c}{\adjustbox{angle=45,lap=\width-0.5em}{#1}}}
\newcolumntype{P}[1]{>{\raggedright\arraybackslash}p{#1}}
\newcommand{\ra}[1]{\renewcommand{\arraystretch}{#1}}
\begin{table}[t]
\centering
\ra{1}
\resizebox{\columnwidth}{!}{%
\begin{tabular}{@{}ld{3.2}d{3.2}d{3.2}d{3.2}d{3.2}@{}}
\textit{Terminating in Country} 
   & \headrow{Brazil}  & \headrow{Netherlands}   & \headrow{India} & \headrow{Kenya} & \headrow{United States}\\
\toprule
Brazil             &.169    &\multicolumn{1}{r}{-}     &\multicolumn{1}{r}{-}    &\multicolumn{1}{r}{-}  &\multicolumn{1}{r}{-} \\ \midrule
Canada             &.001    &.007     &.015      &.006       &\multicolumn{1}{r}{-}  \\
United States      &\cellcolor[HTML]{F7BE81}.774    &\cellcolor[HTML]{F7BE81}.454      &\cellcolor[HTML]{F7BE81}.629      &\cellcolor[HTML]{F7BE81}.443        &\cellcolor[HTML]{F7BE81}.969    \\ \midrule
France             &.001    &.022      &.009      &.023       &.001 \\
Germany            &.002    &.013      &.014      &.028       &.001  \\
Great Britain      &\multicolumn{1}{r}{-}  &.019     &.021     &.032       &.002 \\
Ireland            &.016    &.064      &.027       &.108       &.001   \\
Netherlands        &.013    &\cellcolor[HTML]{F7BE81}.392      &\cellcolor[HTML]{F7BE81}.101      &\cellcolor[HTML]{F7BE81}.200      &.024  \\
Spain              &.001    &\multicolumn{1}{r}{-}     &\multicolumn{1}{r}{-}    &\multicolumn{1}{r}{-}     &\multicolumn{1}{r}{-}    \\ \midrule
Kenya              &\multicolumn{1}{r}{-}        &\multicolumn{1}{r}{-}    &\multicolumn{1}{r}{-}    &.022        &\multicolumn{1}{r}{-}  \\
Mauritius          &\multicolumn{1}{r}{-}      &\multicolumn{1}{r}{-}    &\multicolumn{1}{r}{-}   &.004       &\multicolumn{1}{r}{-}  \\
South Africa       &\multicolumn{1}{r}{-}       &\multicolumn{1}{r}{-}     &\multicolumn{1}{r}{-}  &.021       &\multicolumn{1}{r}{-}  \\ \midrule
United Arab Emirates &\multicolumn{1}{r}{-}     &\multicolumn{1}{r}{-}     &\multicolumn{1}{r}{-}   &.011        &\multicolumn{1}{r}{-}  \\
India              &\multicolumn{1}{r}{-}      &\multicolumn{1}{r}{-}     &.053    &.002        &\multicolumn{1}{r}{-}  \\
Singapore          &\multicolumn{1}{r}{-}       &.002     &\cellcolor[HTML]{F7BE81}.103      &.027       &\multicolumn{1}{r}{-} \\\hline
\end{tabular}
}
\caption{Fraction of paths that terminate in each country by default.}
\label{tab:host}
\end{table}

\begin{table}[t]
\centering
\ra{1}
\resizebox{\columnwidth}{!}{%
\begin{tabular}{@{}ld{3.2}d{3.2}d{3.2}d{3.2}d{3.2}@{}}

\textit{Transiting Country}    & \headrow{Brazil}  & \headrow{Netherlands}   & \headrow{India} & \headrow{Kenya} & \headrow{United States}\\ \toprule
Brazil              &1.00       &\multicolumn{1}{r}{-}   &\multicolumn{1}{r}{-}     &\multicolumn{1}{r}{-}     &\multicolumn{1}{r}{-} \\ \midrule
Canada                &.013       &.007     &.016       &.008      &.081 \\
United States        &\cellcolor[HTML]{F7BE81}.844        &\cellcolor[HTML]{F7BE81}.583     &\cellcolor[HTML]{F7BE81}.715      &\cellcolor[HTML]{F7BE81}.616       &\cellcolor[HTML]{F7BE81}1.00 \\ \midrule
France                 &.059     &.102      &.104       &.221      &.104 \\
Germany                 &.005       &.050    &.032      &.048      &.008 \\
Great Britain                &.024       &\cellcolor[HTML]{F7BE81}.140     &\cellcolor[HTML]{F7BE81}.204      &\cellcolor[HTML]{F7BE81}.500      &.006 \\
Ireland                &.028       &.106      &.031     &.133      &.006 \\
Netherlands                 &.019        &1.00      &.121      &\cellcolor[HTML]{F7BE81}.253      &.031 \\
Spain                  &.176       &.004     &\multicolumn{1}{r}{-}     &\multicolumn{1}{r}{-}      &\multicolumn{1}{r}{-} \\ \midrule
Kenya                 &\multicolumn{1}{r}{-}       &\multicolumn{1}{r}{-}    &\multicolumn{1}{r}{-}      &1.00      &\multicolumn{1}{r}{-} \\
Mauritius                  &\multicolumn{1}{r}{-}       &\multicolumn{1}{r}{-}     &\multicolumn{1}{r}{-}      &\cellcolor[HTML]{F7BE81}.322       &\multicolumn{1}{r}{-} \\
South Africa                 &\multicolumn{1}{r}{-}        &\multicolumn{1}{r}{-}    &\multicolumn{1}{r}{-}     &\cellcolor[HTML]{F7BE81}.334       &\multicolumn{1}{r}{-} \\ \midrule
United Arab Emirates                  &\multicolumn{1}{r}{-}        &\multicolumn{1}{r}{-}    &\multicolumn{1}{r}{-}     &.152       &\multicolumn{1}{r}{-} \\
India               &\multicolumn{1}{r}{-}    &\multicolumn{1}{r}{-}    &1.00     &.058     &\multicolumn{1}{r}{-} \\
Singapore                 &\multicolumn{1}{r}{-}        &.002     &\cellcolor[HTML]{F7BE81}.270       &.040       &.003 \\ \midrule
\end{tabular}
}
\caption{Fraction of paths that each country transits by default.}
\label{tab:transit}
\end{table}

%% file: avoid_results.tex
\section{Preventing Transnational Detours}
\label{avoid_results}

In light of our analysis of the state of default Internet paths from
Section \ref{datasets}, we now explore the extent to which various
techniques and systems can help clients in various countries prevent
unwanted transnational routing detours. We explore two different
mechanisms for increasing path diversity: discovering additional website
replicas by diverting DNS queries through global open DNS resolvers and
creating additional network-layer paths with the use of overlay nodes.
We discuss our measurement methods, develop an avoidance metric and
algorithm, and present our results.

\subsection{Measurement Approach}
\label{avoid_pipelines}

\paragraph{Country Avoidance with Open Resolvers.} If content is
replicated on servers in different parts of the world, open DNS
resolvers located 
around the world may also help clients discover a more diverse set of
replicas.  

\begin{figure}[t]
\centering
\includegraphics[width=.45\textwidth]{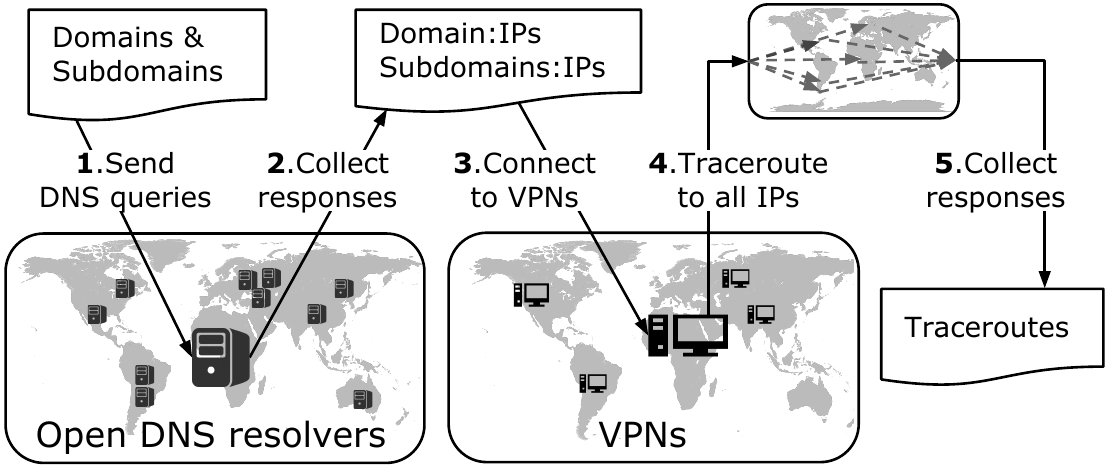}
\caption{Measurement approach for country avoidance with open DNS resolvers.}
\label{fig:avoidance_resolvers}
\end{figure}

Figure~\ref{fig:avoidance_resolvers} illustrates our measurement
approach for this study, which differs slightly from that described in
Section \ref{pipeline}: instead of using RIPE Atlas probes to query
local DNS resolvers, we query open DNS resolvers located around the
world~\cite{open_resolver_list}.  These open DNS resolvers may provide
different IP addresses in the DNS responses, which represent different
locations of content replicas. The measurement study in
Section~\ref{pipeline} used RIPE Atlas probes to traceroute to the IP
addresses in DNS response; in contrast, for this portion of the study we
initiate a VPN connection to the client's country and traceroute
(through the VPN connection) to the IP addresses in the DNS responses
returned by the open resolvers.

\begin{figure}[t]
\centering
\includegraphics[width=.5\textwidth]{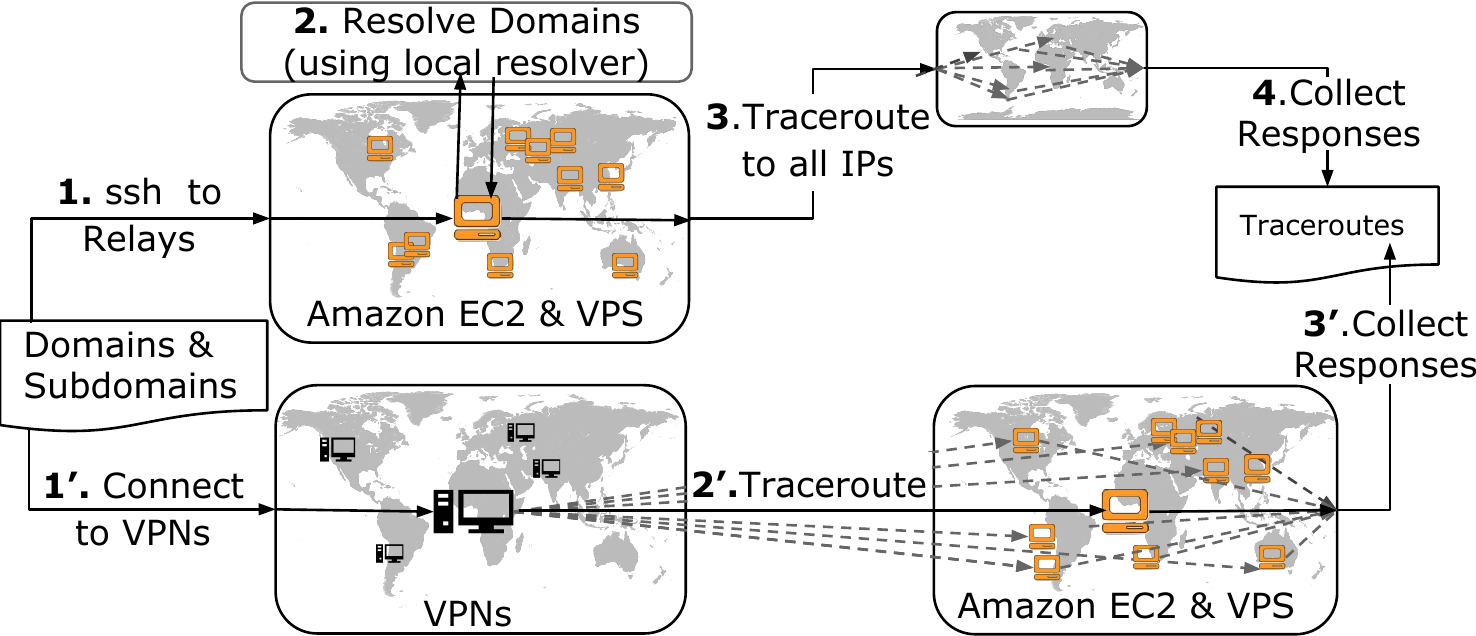}
\caption{Measurement approach for country avoidance with overlay network relays.}
\label{fig:avoidance_relays}
\end{figure}

\paragraph{Country Avoidance with Relays.} Using an overlay network may
help clients route around unfavorable countries or access content that
is hosted in a different country.  Figure~\ref{fig:avoidance_relays}
shows the steps to conduct this measurement. 
After selecting relay machines, we run traceroute measurements from
Country X to each relay and from each relay to the set of domains. We
then analyze these traceroutes using the pipeline in Figure
\ref{fig:analysis_pipeline} to determine country-level paths. 

We use eight Amazon EC2 instances, one in each geographic region (United States, Ireland, Germany, Singapore, South Korea, Japan, Australia, Brazil), as well as 4 Virtual Private Server (VPS) machines (France, Spain, Brazil, Singapore), which are virtual machines that are functionally equivalent to dedicated physical servers.  The conjunction of these two sets of machines allow us to evaluate surveillance avoidance with a geographically diverse set of relays. By selecting an open resolver in each country that also has a relay in it we can keep the variation in measurement methods low, leading to a more accurate comparison of country avoidance methods.

\subsection{Avoidability Metrics}
\label{metrics}

We introduce a new metric and algorithm to measure how often a client in
Country X can avoid a specific country Y.  Using the proposed metric
and algorithm, we can compare how well the different methods achieve
country avoidance for any (X, Y) pair.

\paragraph{Avoidability metric.}  We introduce an avoidability metric to
quantify how often
traffic can avoid Country Y when it originates in Country X.
Avoidability is the fraction of paths that start in Country
X and do not transit Country Y.  We calculate this value by dividing the
number of paths from Country X to domains that do not traverse Country Y
by the total number of paths from Country X. The resulting value will be
in the range [0,1], where 0 means the country is unavoidable for all of
the domains in our study, and 1 means the client can avoid Country Y for
all domains in our study.  For example, there are three paths
originating in Brazil: (1)~$BR \rightarrow US$, (2)~$BR \rightarrow CO
\rightarrow None$, 3) $BR \rightarrow *** \rightarrow BR$.  After
processing the paths as described in Section \ref{c_map}, the resulting
paths are: (1)~$BR \rightarrow US$, (2)~$BR \rightarrow CO$, (3)~$BR
\rightarrow BR$.  The avoidance value for avoiding the United States
would be 2/3 because two out of the three paths do not traverse the
United States.  This metric represents a lower bound,
because it is possible that the third path timed out ($***$) because it
traversed the United States, which would make the third path: $BR
\rightarrow US \rightarrow BR$, and would cause the avoidance metric to
drop to 1/3.

\paragraph{Avoidability algorithm with open resolvers.} Recall from the measurement pipeline for avoidance with open resolvers, described in Section \ref{avoid_pipelines}, that the resulting data are traceroutes from the client in Country X to \textit{all} IP addresses in \textit{all} open DNS resolver responses.  To measure avoidability, there must exist at least one path from the client in Country X to the domain for the client to be able to avoid Country Y when accessing the domain.  The country avoidance value is the fraction of domains accessible from the client in Country X without traversing Country Y.

\paragraph{Avoidability algorithm with relays.}  Measuring the avoidability of a Country Y from a client in Country X using relays has two components: (1)~Is Country Y on the path from the client in Country X to the relay?  (2)~Is Country Y on the path from the relay to the domain?  For every domain, our algorithm checks if there exists at least one path from the client in Country X through any relay and on to the domain, and does not transit Country Y.  
The algorithm (Algorithm~\ref{avoid_algo}) produces a value in the range
[0,1] that can be compared to the output of the avoidability metric
described above.   

\begin{algorithm}[t]
\caption{Avoidability Algorithm}
\label{avoid_algo}
\small
\begin{algorithmic}[1]
\Function{CalcAvoidance}{set $paths1$, set $paths2$, string c}
    \State set $usableRelays$
    \For{each $(relay,path)$ in $paths1$} 
    	\If{$c$ not in $path$}
		\State $usableRelays \gets path$
	\EndIf
    \EndFor
    \State set $accessibleDomains$
    \For{each $(relay,domain,path$ in $paths2$}
    \If{$relay$ in $usableRelays$}
        \If{$c$ not in $path$}
        \State $accessibleDomains \gets domain$
        \EndIf
    \EndIf
    \EndFor
    \State $D \gets$ number of all unique domains in $paths2$
    \State $A \gets$ length of $accessibleDomains$
    \State \Return $A / D$
\EndFunction
\end{algorithmic}
\end{algorithm}

\paragraph{Upper bound on avoidability.}  Although the avoidability
metric and algorithm provide a method to quantify how avoidable Country
Y is from a client in Country X, it may be the case that a number of
domains are only hosted in Country Y, so the avoidance value for these
countries would never reach 1.0.  For this reason, we measured the {\em upper
bound} on avoidance for a given pair of (Country X, Country Y) that
represents the best case value for avoidance.  
Algorithm \ref{upperbound_algo} shows the pseudocode for computing this metric. The algorithm analyzes the destinations of all domains from all relays and if there exists at least one destination for a domain that is not in Country Y, then this increases the upper bound value.  An upper bound value of 1.0 means that every domain studied is hosted (or has a replica) outside of Country Y.  This value puts the avoidance values in perspective for each (Country X, Country Y) pair. 

\begin{algorithm}[t]
\caption{Avoidance Upper Bound Algorithm}
\label{upperbound_algo}
\small
\begin{algorithmic}[1]
\Function{CalcUpperbound}{set $relayDomainPaths$, string $c$}
    \State $zeros(domainLocations)$
    \For{each $(r,d,p)$ in $relayDomainPaths$} 
		\State $dest \gets $ last item in $p$
		\State $domainLocations[d] \gets dest$
    \EndFor
    \State set $accessibleDomains$
    \For{each $domain$ in $domainLocations$}
    \If{$domainLocations[domain] \neq $ set[$c$]}
    \State $accessibleDomains \gets domain$
    \EndIf
    \EndFor
    \State $D \gets$ all unique domains in  $relayDomainPaths$
    \State $A \gets$ length of $accessibleDomains$
    \State \Return $A / D$
\EndFunction
\end{algorithmic}
\end{algorithm}

\subsection{Results}

We compared avoidance values when using open resolvers, when using
relays, and when using no country avoidance tool.  First, we discuss how
effective open resolvers are at country avoidance.  We then examine the
effectiveness of relays for country avoidance, as well as for keeping local
traffic local.  Table \ref{tab:avoid} shows avoidance values; the top
row shows the countries we studied and the left column shows the country
that the client aims to avoid.

\newcolumntype{d}[1]{D{.}{.}{#1}}
\begin{table*}[t]
\centering
\resizebox{\textwidth}{!}{%
\begin{tabular}{P{32mm}|d{3.2}d{3.2}d{3.2}|d{3.2}d{3.2}d{3.2}|d{3.2}d{3.2}d{3.2}|d{3.2}d{3.2}d{3.2}|d{3.2}d{3.2}d{3.2}}
\multicolumn{1}{l}{}    & \headrow{No Relay} &\headrow{Open Resolvers} & \headrow{Relays} & \headrow{No Relay} &\headrow{Open Resolvers} & \headrow{Relays} & \headrow{No Relay} &\headrow{Open Resolvers} & \headrow{Relays}   & \headrow{No Relay} &\headrow{Open Resolvers} & \headrow{Relays}  & \headrow{No Relay} &\headrow{Open Resolvers} & \headrow{Relays} \\ \toprule
\textit{Country to Avoid}    &\multicolumn{3}{c|}{\textit{Brazil}}   &\multicolumn{3}{c|}{\textit{Netherlands}}   &\multicolumn{3}{c|}{\textit{India}} &\multicolumn{3}{c|}{\textit{Kenya}} &\multicolumn{3}{c}{\textit{United States}}\\ \toprule

Brazil               &0.00  &0.00   &0.00     &1.00 &1.00 &1.00   &1.00  &1.00  &1.00   &1.00 &1.00  &1.00  &1.00 &1.00 &1.00  \\ \midrule
Canada               &.98  &1.00   &1.00     &.99 &1.00 &1.00   &.98  &.98  &.98  &.99 &.99  &.99  &.92 &1.00  &1.00  \\
United States        &\cellcolor[HTML]{F7BE81}.15  &\cellcolor[HTML]{F7BE81}.19   &\cellcolor[HTML]{F7BE81}.62     &\cellcolor[HTML]{F7BE81}.41  &\cellcolor[HTML]{F7BE81}.57 &\cellcolor[HTML]{F7BE81}.63   &\cellcolor[HTML]{F7BE81}.28  &\cellcolor[HTML]{F7BE81}.45  &\cellcolor[HTML]{F7BE81}.65  &\cellcolor[HTML]{F7BE81}.38 &\cellcolor[HTML]{F7BE81}.55  &\cellcolor[HTML]{F7BE81}.40  &\cellcolor[HTML]{F7BE81}0.00  &\cellcolor[HTML]{F7BE81}0.00 &\cellcolor[HTML]{F7BE81}0.00  \\ \midrule
France               &.94   &.98  &1.00     &.89 &.96 &.99   &.89  &.98  &1.00  &.77 &.89  &.98  &.89 &.99 &.99  \\
Germany              &.99  &.99   &1.00     &.95 &.98 &.99   &.96  &.97  &.99  &.95 &.99  &1.00  &.99 &.99 &1.00  \\
Great Britain        &.97  &.97   &1.00     &.86 &.87 &.99   &\cellcolor[HTML]{F7BE81}.79  &\cellcolor[HTML]{F7BE81}.79  &\cellcolor[HTML]{F7BE81}1.00  &\cellcolor[HTML]{F7BE81}.50 &\cellcolor[HTML]{F7BE81}.71  &\cellcolor[HTML]{F7BE81}.97  &.99 &.99 &1.00  \\
Ireland              &.97   &.98  &.99     &.89 &.97 &.99   &.96 &.99   &.99  &.86 &.98  &.99  &.99 &.99 &.99  \\
Netherlands          &.98  &.98   &.99     &0.00 &0.00 &0.00   &.87  &.98  &.99  &\cellcolor[HTML]{F7BE81}.74  &\cellcolor[HTML]{F7BE81}.98 &\cellcolor[HTML]{F7BE81}.99  &.97 &.99 &.99  \\
Spain                &.82  &1.00   &1.00     &.99 &.99 &.99   &1.00  &.99  &1.00  &1.00  &1.00 &1.00  &1.00 &1.00  &1.00  \\ \midrule
Kenya                &1.00  &1.00   &1.00     &1.00 &1.00 &1.00   &1.00  &1.00  &1.00  &0.00 &0.00  &0.00  &1.00 &1.00  &1.00  \\
Mauritius            &1.00  &1.00   &1.00     &1.00 &1.00 &1.00   &1.00  &1.00  &1.00  &\cellcolor[HTML]{F7BE81}.67 &\cellcolor[HTML]{F7BE81}.97  &\cellcolor[HTML]{F7BE81}.99  &1.00 &1.00 &1.00  \\
South Africa         &1.00  &1.00   &1.00     &1.00 &1.00 &1.00   &1.00 &1.00   &1.00  &\cellcolor[HTML]{F7BE81}.66 &\cellcolor[HTML]{F7BE81}.87  &\cellcolor[HTML]{F7BE81}.66  &1.00 &1.00 &1.00  \\ \midrule
United Arab Emirates &1.00  &1.00   &1.00     &1.00 &1.00 &1.00   &1.00 &1.00   &1.00  &\cellcolor[HTML]{F7BE81}.84 &\cellcolor[HTML]{F7BE81}1.00  &\cellcolor[HTML]{F7BE81}.99  &1.00 &1.00 &1.00  \\
India                &1.00  &1.00   &1.00     &.99 &1.00 &1.00   &0.00 &0.00   &0.00  &.94 &.94  &1.00  &.99 &1.00 &1.00  \\
Singapore            &.99  &.99   &1.00     &.99 &.99 &1.00   &\cellcolor[HTML]{F7BE81}.73  &\cellcolor[HTML]{F7BE81}.92  &\cellcolor[HTML]{F7BE81}.94  &.96 &.96  &1.00  &.99 &.99 &1.00  \\\midrule
\end{tabular}
}
\caption{Avoidance values for different techniques of country avoidance.  The upper bound on avoidance is 1.0 in most cases, but not all.  It is 
common for some European countries to host a domain, and therefore the upper bound is slightly lower than 1.0.  The upper bound on avoidance of the 
United States is significantly lower than the upper bound on avoidance for any other country; .886, .790, .844, and .765 are the upper bounds on avoidance 
of the United States for paths originating in Brazil, Netherlands, India, and Kenya, respectively.}
\label{tab:avoid}
\end{table*}

\subsubsection{Avoidance with Open Resolvers}

A given country is more avoidable (higher avoidance value) when open
resolvers are used as a tool for country avoidance. 

\begin{finding}[Open Resolver Effectiveness]
Using open DNS resolvers for country avoidance achieves more country
avoidance than using local resolvers and less (or equal)
avoidance than using relays for clients in most
countries. 
\end{finding}
\noindent
For Brazilian paths, open resolvers only achieve 4\% more avoidance than
using local resolvers when avoiding the United States, whereas relays
achieve 47\% more avoidance.  On the other hand, open resolvers are
about as effective as relays are for avoidance for paths originating in
the United States.   

\begin{finding}[Kenya as an Outlier]
For clients in Kenya, open DNS resolvers are significantly more
effective than relays for 
avoiding the United States, South Africa, and the United Arab Emirates.
\end{finding}
\noindent
Clients in Kenya should use open DNS resolvers when avoiding specific
countries, as they can avoid these specific countries more often than
when using relays.  Kenyan clients can avoid the United States for 55\%
of paths when using open resolvers, whereas they can only avoid the United States for 40\% of
paths when using relays.  The difference in how often the United States can be avoided can be
attributed to the lower amount of DNS diversity when using relays as
compared to using open resolvers. For a client in Kenya trying to avoid
the United States, the client can only use the relay located in Ireland
(because all paths from the client to the other relays traverse the
United States), and therefore only gets DNS responses from locally
resolving domains on the Ireland relay.  When using open resolvers, the
client gets more DNS diversity as he gets DNS responses from all open
resolvers located in different countries. 

The amount of avoidance Kenyan clients can achieve for avoiding South Africa is the same,
regardless of whether the client is using relays, because all paths
between the client and the relays traverse South Africa.  Fortunately,
clients can avoid South Africa for significantly more paths when using
open resolvers, likely as a result of the fact that open DNS resolvers
can better uncover hosting diversity.   

\subsubsection{Avoidance with Relays}
As seen in Table \ref{tab:avoid}, there are two significant trends: 1) the ability for a client to avoid a given Country Y increases with the use of relays, and 2) the least avoidable countries are surveillance states.

\begin{finding}[Relay Effectiveness]
For 84\% of the (Country X, Country Y) pairs shown in Table \ref{tab:avoid} the avoidance with relays reaches the upper bound on avoidance. 
\end{finding}
\noindent
In almost every (Country X, Country Y) pair, where Country X is the
client's country (Brazil, Netherlands, India, Kenya, or the United
States) and Country Y is the country to avoid, the use of an overlay
network makes Country Y more avoidable than the default routes.  The one
exception we encountered is when a client is located in Kenya and wants
to avoid South Africa, where, as mentioned, all paths through our
relays exit Kenya via South Africa.

\begin{finding}[Relays Achieve Upper Bound]
Clients in the United States can achieve the upper bound of avoidance for all countries---relays help clients in this country avoid all other Country Y in all cases that the domain is not hosted in Country~Y.  
\end{finding}
\noindent
Relays are most effective for clients in the United States.  On the other hand, it is much rarer for (Kenya, Country Y) pairs to achieve the upper bound of surveillance, showing that it is more difficult for Kenyan clients to avoid a given country.  This is not to say that relays are not effective for clients in Kenya; for example, the default routes to the top 100 domains for Kenyans avoid Great Britain 50\% of the time, but with relays this percentage increases to about 97\% of the time, and the upper bound is about 98\%. 

\begin{finding}[Surveillance States are Less Avoidable]
The ability for any country to avoid the United States is significantly lower than it's ability to avoid any other country in all four situations: without relays, with open resolvers, with relays, and the upper bound. 
\end{finding}
\noindent
Despite increasing clients' ability to avoid the United States, relays are not as effective at helping clients avoid this country as compared to the effectiveness of the relays at avoiding all other Country Y.  Clients in India can avoid the United States more often than clients in Brazil, Netherlands, and Kenya, by avoiding the United States for 65\% of paths.  Kenyan clients can only avoid the United States 40\% of the time even while using relays.  Additionally, the upper bound for avoiding the United States is significantly lower in comparison to any other country.  

\begin{finding}[Keeping Local Traffic Local]
Using relays decreased both the number of tromboning paths, and the
number of countries involved in tromboning paths.
\end{finding}
\noindent
For the cases where there were relays located in one of the five studied countries, we evaluated how effectively the use of relays kept local traffic local.  This evaluation was possible for Brazil and the United States.  Tromboning Brazilian paths decreased from 13.2\% without relays to 9.7\% with relays; when relays are used, all tromboning paths goes only to the United States.  With the use of relays, there was only 1.3\% tromboning paths for a United States client, whereas without relays there was 11.2\% tromboning paths.  For the 1.2\% of paths that trombones from the United States, it goes only to Ireland.

%% file: discussion.tex
\section{Discussion}
\label{discussion}

\paragraph{Avoiding multiple countries.} 
We have studied only the extent to which Internet paths can be
engineered to avoid a {single} country.  Yet, avoiding a single country
may force an Internet path into {\em other} unfavorable
jurisdictions. This possibility suggests that we should also be
exploring the feasibility of avoiding multiple surveillance states (\eg,
the ``Five Eyes'') or perhaps even entire regions. It is already clear
that avoiding certain combinations of countries is not possible, at
least given the current set of relays; for
example, to avoid the US, Kenyan clients rely on the relay located in
Ireland, so avoiding both countries is often impossible.

\paragraph{The evolution of routing detours and avoidance over time.}
Our study is based on a snapshot of Internet paths. Over time, paths
change, hosting locations change, IXPs are built, submarine cables are
laid, and surveillance states change.  Future work can and should
involve exploring how these paths evolve over time, and analyzing the
relative effectiveness of different strategies for controlling traffic flows.

\paragraph{Isolating DNS diversity vs. path diversity.}
In our experiments, the overlay network relays perform DNS lookups from
geographically diverse locations, which provides some level of DNS
diversity in addition to the path diversity that the relays inherently
provide. This approach somewhat conflates the benefits of DNS diversity
with the benefits of path diversity and in practice may increase
clients' vulnerability to surveillance, since each relay is performing
DNS lookups on each client's behalf. We plan to conduct additional
experiments where the client relies on its local DNS resolver to map
domains to IP addresses, as opposed to relying on the relays for both
DNS resolution and routing diversity.

%% file: related_work.tex
\section{Related Work}
\label{related}

\paragraph{Nation-state routing analysis.}  Recently, Shah and
Papadopoulos measured international BGP detours (paths that originate in
one country, cross international borders, and then return to the
original country)~\cite{shah2015characterizing}.  Using BGP routing
tables, they found 2 million detours in each month of their study (out
of 7 billion total paths), and they then characterized the detours based
on detour dynamics and persistence.  Our work differs by actively
measuring traceroutes (actual paths), as opposed to analyzing BGP
routes.  This difference is fundamental as BGP provides the AS path
announced in BGP update messages, which is not necessarily the same as
the actual path of data packets.  Obar and Clement analyzed traceroutes
that started and ended in Canada, but ``boomeranged'' through the United
States (``boomerang'' is another term for tromboning), and argued that
this is a violation of Canadian network
sovereignty~\cite{obar2012internet}.  Most closely related to our work,
Karlin et al. developed a framework for country-level
routing analysis to study how much influence each country has over
interdomain routing~\cite{karlin2009nation}.  This work measures the
centrality of a country to routing and uses AS-path inference to measure
and quantify country centrality, whereas our work uses active
measurements and measures avoidability of a given country. 

\paragraph{Mapping national Internet topologies.}  In 2011, Roberts et al. described
a method for mapping national networks of ASes, identifying ASes that
act as points of control in the national network, and measuring the
complexity of the national network~\cite{roberts2011mapping}.  There
have also been a number of studies that measured and classified the
network within a country.  Wahlisch et al. measured and classified the
ASes on the German Internet~\cite{wahlisch2010framework,
  wahlisch2012exposing}, Zhou et al. measured the complete
Chinese Internet topology at the AS level~\cite{zhou2007chinese}, and
Bischof et al. characterized the current state of Cuba's
connectivity with the rest of the world~\cite{bischof2015and}.
Interconnectivity has also been studied at the continent level; Gupta
et al. first looked at ISP interconnectivity within
Africa~\cite{gupta2014peering}, and it was studied later by Fanou et al.~\cite{fanou2015diversity}.

\paragraph{Circumvention Systems.}  There has been research into
circumvention systems, particularly for censorship circumvention, that
is related this work, but not sufficient for surveillance circumvention.
Tor is an anonymity system that uses three relays and layered encryption
to allow users to communicate anonymously~\cite{dingledine2004tor}.
VPNGate is a public VPN relay system aimed at circumventing national
firewalls~\cite{nobori2014vpn}.  Unfortunately, VPNGate does not allow a
client to choose any available VPN, which makes surveillance avoidance
harder.

%% file: conclusion.tex
\section{Conclusion}
\label{conclusion}

We have measured Internet paths to characterize routing
detours that take Internet paths through countries that perform
surveillance.  Our findings show that paths commonly traverse known
surveillance states, even when they originate and end in a
non-surveillance state.  As a possible step towards a remedy, we have
investigated how clients can use the open DNS resolver infrastructure
and overlay network relays to prevent routing detours through
unfavorable jurisdictions.  These methods give clients the power to
avoid certain countries, as well as help keep local traffic local.
Although some countries are completely avoidable, we find that some of
the more prominent surveillance states are the least avoidable.

Our study presents several opportunities for follow-up studies and
future work. First, Internet paths continually
evolve; we will repeat this analysis over time and publish the results
and data on a public website, to help deepen our collective
understanding about how the evolution of Internet connectivity affects
transnational routes. Second, our analysis should be extended to study
the extent to which citizens in one country can avoid groups of
countries or even entire regions. Finally, although our results provide strong 
evidence for the existence of various transnational data flows, factors
such as uncertain IP geolocation make it difficult to provide clients
guarantees about country-level avoidance; developing techniques and
systems that offer clients stronger guarantees
is a ripe opportunity for future work.